\documentclass[journal,twocolumn]{IEEEtran}
\usepackage{lipsum}
\usepackage{url}
\usepackage{mathtools}
\usepackage{amsmath,amssymb}
 \usepackage{graphicx}
 \usepackage{booktabs}
  \usepackage[dvipsnames]{xcolor}
  \usepackage{multirow}

\ifCLASSINFOpdf
\else
\fi
\begin{document}
%
%
%
%
\title{BEHM-GAN: Bandwidth Extension of Historical Music using Generative Adversarial Networks}

\author{Eloi Moliner and Vesa V\"alim\"aki, \IEEEmembership{Fellow, IEEE}

\thanks{Manuscript received February 16, 2022; revised June 15, 2022. This research is part of the activities of the Nordic Sound and Music Computing Network---NordicSMC, NordForsk project no.~86892. \emph{(Corresponding author: Eloi Moliner)}}
\thanks{E. Moliner and V. Välimäki are with the Acoustics Laboratory, Department of Signal Processing and Acoustics, Aalto University, Espoo, Finland (e-mail: eloi.moliner@aalto.fi).}
}

%
%

\markboth{SUBMITTED TO IEEE/ACM TRANSACTIONS ON AUDIO, SPEECH, AND LANGUAGE PROCESSING, 2022}%
{Shell \MakeLowercase{\textit{et al.}}: Some Title about Bandwidth Extension}
%



\maketitle

\begin{abstract}
Audio bandwidth extension aims to expand the spectrum of bandlimited audio signals. Although this topic has been broadly studied during recent years, the particular problem of extending the bandwidth of historical music recordings remains an open challenge. This paper proposes a method for the bandwidth extension of historical music using generative adversarial networks (BEHM-GAN) as a practical solution to this problem. The proposed method works with the complex spectrogram representation of audio and, thanks to a dedicated regularization strategy, can effectively extend the bandwidth of out-of-distribution real historical recordings. The BEHM-GAN is designed to be applied as a second step after denoising the recording to suppress any additive disturbances, such as clicks and background noise. We train and evaluate the method using solo piano classical music. The proposed method outperforms the compared baselines in both objective and subjective experiments. The results of a formal blind listening test show that BEHM-GAN significantly increases the perceptual sound quality in early-20th-century gramophone recordings. For several items, there is a substantial improvement in the mean opinion score after enhancing historical recordings with the proposed bandwidth-extension algorithm. This study represents a relevant step toward data-driven music restoration in real-world scenarios.

\end{abstract}
\begin{IEEEkeywords}
Audio recording, convolutional neural networks, machine learning, music, signal restoration. 
\end{IEEEkeywords}

%
\IEEEpeerreviewmaketitle

\section{Introduction}
%
%
%
%


\IEEEPARstart{H}{istorical} music recordings are available in large numbers in archives but, due to the technological limitations of the time, by modern standard they  are of a very poor audio quality. 
Early-20th-century gramophone recordings suffer from severe degradations, such as multiple kinds of surface noises, distortion, and a narrow frequency bandwidth \cite{godsill_digital_1998, Esquef2008}. 
The goal of digital audio restoration is to correct the imperfections of audio recordings so that the resulting sound quality is enhanced. Restoration may target the removal of clicks and noises \cite{rund_evaluation_2021, li2020learning}, the inpainting of missing audio segments \cite{perraudin2018inpainting, marafioti_gacela_2020}, declipping \cite{zavivska2020survey}, or the bandwidth extension of bandlimited audio signals, among other tasks.

This paper focuses on audio bandwidth extension and, particularly, on applying it to historical music recordings. During recent years, many works have used modern deep learning technologies for bandwidth extension. However, their goal has usually been to enhance modern digital audio signals having a limited bandwidth because of the usage of a lower sampling rate. Only a few exceptions are relevant to music signal processing \cite{miron2018high, lagrange2020bandwidth, hu_phase-aware_2020}, whereas most of these studies focus on processing speech \cite{kuleshov2017audio, Wang2021, su_bandwidth_2021,li_real-time_2021,lee_nu-wave_2021}. Although music and speech share the same domain of acoustic signals, the two are fundamentally different.

Usually, the aforementioned methods are trained in a self-supervised fashion by pre-processing the audio data with lowpass filters to simulate the bandwidth limitation. Then, the models are optimized to extend the input lowpass-filtered audio using the broadband original signal as a target. However, bandwidth-extending historical recordings entails an extra challenge, as no full-bandwidth version is available for this particular material. Then, we would rely on the model, trained with synthetically filtered data, to extrapolate to real historical recordings, a harder out-of-distribution scenario. One should also consider the problem of filter generalization \cite{sulun_filter_2020}, which refers to the inability deep neural networks to generalize when they are trained using a single type of lowpass filter in the training-data pipeline. 

\begin{figure}
    \centering
    \includegraphics{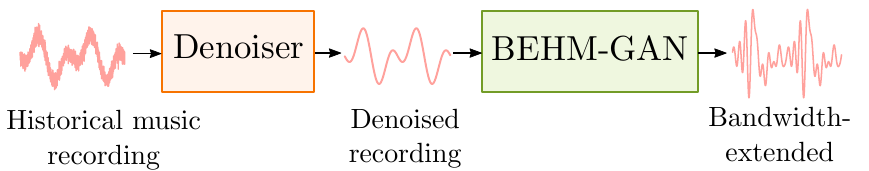}
    \caption{Illustration of the inference pipeline. The denoiser block refers to a  model borrowed from earlier work \cite{Moliner2022}, and the BEHM-GAN is the model proposed in this paper.}
    \label{inference}
\end{figure}

Another problem with old gramophone recordings is that they are often corrupted with a wide range of global and local disturbances, such as hiss, clicks, and thumps. These additive noises represent another obstacle in enhancing the recording. Luckily, recent works have shown that a vast majority of clicks and noises appearing on gramophone recordings can be efficiently suppressed using deep-learning models  \cite{li2020learning, Moliner2022}. We studied this problem in particular and proposed a model consisting of a spectrogram-based deep-neural-network architecture \cite{Moliner2022}. The denoising model was trained using a realistic dataset of noise samples extracted from gramophone recordings and yields a considerable enhancement in quality \cite{Moliner2022}. This paper builds upon this previous work \cite{Moliner2022}, in such a way that the proposed bandwidth extension method is intended to be applied as a second step after the original recording has been first denoised, as illustrated in Fig.~\ref{inference}.

In this paper, we present a method for the bandwidth extension of historical music recordings using generative adversarial networks (BEHM-GAN) and evaluate it with solo piano music recordings. The proposed method is based on a generative adversarial network (GAN) \cite{goodfellow2014generative} and combines a generator in the spectrogram domain with multiple time-domain discriminators. To provide the model with the necessary robustness to make inference in historical recordings, we propose a simple but effective filter regularization in the training stage. The strategy is based on adding a small amount of white Gaussian noise after applying a lowpass filter with a randomized cutoff frequency. We show that the proposed method inserts sound energy in an appropriate way above the cutoff frequency of about 3\,kHz and, as determined in a formal blind listening test, significantly improves the perceived quality of both artificially bandlimited and real old piano music. As far as we are aware, this is the first work that successfully extends the bandwidth in real historical music recordings. We emphasize that the goal of this work is not to restore exactly the missing sound events, but to recreate plausible high-frequency content and, thus, make the music more pleasant to listen to.

The remainder of this paper is organized as follows. Sec.~\ref{sec:relatedwork} reviews the most relevant related work, with a focus on recent deep-learning-based audio-bandwidth-extension studies. To understand the bandwidth limitation of historical gramophone recordings, Sec.~\ref{sec:Spectral_Analysis} analyzes empirically their spectral characteristics using the Long-Term Average Spectrum (LTAS). The proposed BEHM-GAN method is introduced in Sec. \ref{sec:methods} and is evaluated in Sec.~\ref{sec:experiments} with objective and subjective metrics. In order to assess the robustness of the compared methods, the experiments are conducted under two separate conditions: lowpass filtered modern recordings and old historical recordings. Finally, Sec.~\ref{sec:conclusion} concludes the paper.

\section{Related Work}
\label{sec:relatedwork}
This section reviews previous work on audio bandwidth extension, focusing on approaches that apply GANs for related tasks or study the problem of filter generalization.

\subsection{Audio Bandwidth Extension}

Audio bandwidth extension refers to methods that extend the  spectrum of audio signals \cite{larsen2005audio}. A popular sub-topic is audio super-resolution \cite{kuleshov2017audio, Wang2021}, which increases the sampling rate of a given audio signal by extending its bandwidth above the original Nyquist limit. This topic has a long history in telephony, where the bandwidth of a transmitted speech signal was usually compressed because of channel constraints \cite{nilsson2001avoiding}. Another relevant application is audio compression, as bandwidth-extension techniques can be used to reduce the bit rate of an audio signal \cite{ziegler2002enhancing, huang2019a}.

Early works used signal processing methods such as a source-filter model \cite{Makhoul1979High, Johannes2016Asubjective}, codebook mapping \cite{Carl1994bandwidth}, nonlinear devices \cite{larsen2005audio}, or spectral band replication \cite{dietz2002spectral}. Other approaches were based on data-driven techniques, such as Gaussian mixture models \cite{Park2000Narrowband,SeoHyunson}, hidden Markov models \cite{Jax}, or shallow neural networks \cite{Kontio2007, pulakka2011bandwidth}. However, due to their inadequate modeling capabilities, these early methods often lead to a poor or mediocre audio quality.


More recently,  deep-learning-based bandwidth extension-methods outperformed previous approaches. The vast majority of the presented methods used convolutional neural networks and work using either a spectrogram representation \cite{Li2015},\cite{DBLP:conf/interspeech/LiHXL15}, raw audio data \cite{kuleshov2017audio,Gupta2019Speech, birnbaum2021temporal, Wang2021}, or a mixture of both \cite{ yian_lim_time-frequency_2018, lin_two-stage_2021}.

\subsection{GANs for Audio Bandwidth Extension}

Deep learning models based on optimizing reconstruction losses excel at tasks where the goal is to design a nonlinear mapping between two data distributions, e.g., denoising. However, the performance of supervised learning is limited when the task involves generating new content that is absent in the observed signal, as is the case in bandwidth extension. As a consequence, deep learning tends to build over-smoothed and unrealistic spectra. For this reason, recent works have adopted a generative approach that allows the model to have more expressive power. Some studies applied different kinds of generative models for the task of speech bandwidth extension, such as flow-based models  \cite{zhang_wsrglow_2021} or diffusion-probabilistic models \cite{lee_nu-wave_2021}, but, in particular, GANs \cite{goodfellow2014generative} have shown great potential for this task. 

GANs are generative models that are based on optimizing a two-player min-max game between a generator $G$ and a discriminator $D$ \cite{goodfellow2014generative}. The discriminator $D$ is optimized to distinguish real data samples from the ones generated by $G$, whereas $G$ tries to fool $D$ by generating data samples that are harder to detect. Ideally, if the training does not collapse, both $G$ and $D$ will converge to the so-called Nash equilibrium, where $G$ fits the target data distribution and $D$ is unable to detect the fake data samples from the real ones. In the original GAN formulation, a latent vector of Gaussian noise $z$ is provided to the generator $G(z)$ as an input. However, GANs for audio bandwidth extension can be viewed as conditional GANs \cite{mirza2014conditional}, where the generator $G(x,z)$ is also conditioned on an observed signal $x$, here the bandlimited input. Then, due to the high dimensionality of $x$, the latent vector $z$ is often omitted if a controllable latent-space representation is not required \cite{isola2017image}.

Although only a few studies have applied GAN models to bandwidth extension of music signals \cite{hu_phase-aware_2020, kim_bandwidth_2021}, many recent works have applied them for speech \cite{li_real-time_2021, su_bandwidth_2021,Eskimez2019Adversarial}. Eskimez \emph{et al.} \cite{Eskimez2019Adversarial} proposed one of the earliest works using an adversarial approach for speech super-resolution. Their proposed model predicted the magnitude spectrogram representation of audio and used an adversarial loss combined with a reconstruction loss. However, the Eskimez model had the limitation that it did not predict the phase information but just replicated it \cite{Eskimez2019Adversarial}. Other phase-aware works made an effort to incorporate the phase information into the training framework \cite{hu_phase-aware_2020}. Instead, Kim \emph{et al.} \cite{kim_bandwidth_2021}, opted for working directly on raw audio, thus avoiding the aforementioned phase issues. They also incorporated a third auxiliary feature-matching loss term. Su \emph{et al.} \cite{su_bandwidth_2021} used a time-domain Wavenet generator and a composite of multiple time-domain and spectral-domain discriminators. Utilizing a complex combination of loss terms, they achieved impressive results. Li \emph{et al.} \cite{li_real-time_2021} proposed a lighter time-domain model that was suitable to run in real-time.

\subsection{Lowpass Filter Generalization}
\label{sec:lowpass}

A particular problem in the audio-bandwidth-extension literature is the incapability of deep neural networks to generalize when they are trained using lowpass filters. We hypothesize that this problem is a special case of shortcut learning \cite{geirhos2020shortcut}, stating that the model does not learn the true underlying mechanisms of the data but relies on spurious statistical relationships. In this case, the neural network learns the easier task of inverting the response of a lowpass filter instead of generating new and coherent high-frequency content.
 
Kuleshov \emph{et al}. \cite{kuleshov2017audio} observed that a neural network trained to conduct audio super-resolution using aliased training data was ineffective if an antialiasing filter was included during testing. The same problem happened when antialising filters were only used during training and when the filters utilized during training and testing differed. Sulun and Davies \cite{sulun_filter_2020} studied this phenomenon and named it "filter overfitting". They showed that the problem could be mitigated considerably by using a set of different lowpass filters during training as a data augmentation strategy. 
 
Wang and Wang \cite{Wang2021} examined the robustness of their speech super-resolution model that was trained with different down-sampling schemes. They proposed a solution based on randomly combining three different down-sampling strategies during training. Li \emph{et al}. \cite{li_real-time_2021} also experimented with using variable-band filters with randomized cutoff frequencies to increase the robustness of the model in real-life speech bandwidth-extension scenarios. Similarly, Nguyen \emph{et al}. \cite{nguyen2021tunet} applied anti-aliasing filters having random order and ripple intending to improve the robustness of their model.

 This problem gets more relevant in the case of historical recordings when we aim to infer a target distribution that has not been processed by any lowpass filter. In this case, neither a specific filter specification nor a known cutoff frequency can be assumed, since they may vary greatly depending on the recording conditions. In the next section, we investigate the underlying lowpass filtering in old gramophone recordings.


%
\section{Spectral Analysis of Gramophone Recordings}
\label{sec:Spectral_Analysis}

 \begin{figure}[t]
    \centering
    \includegraphics{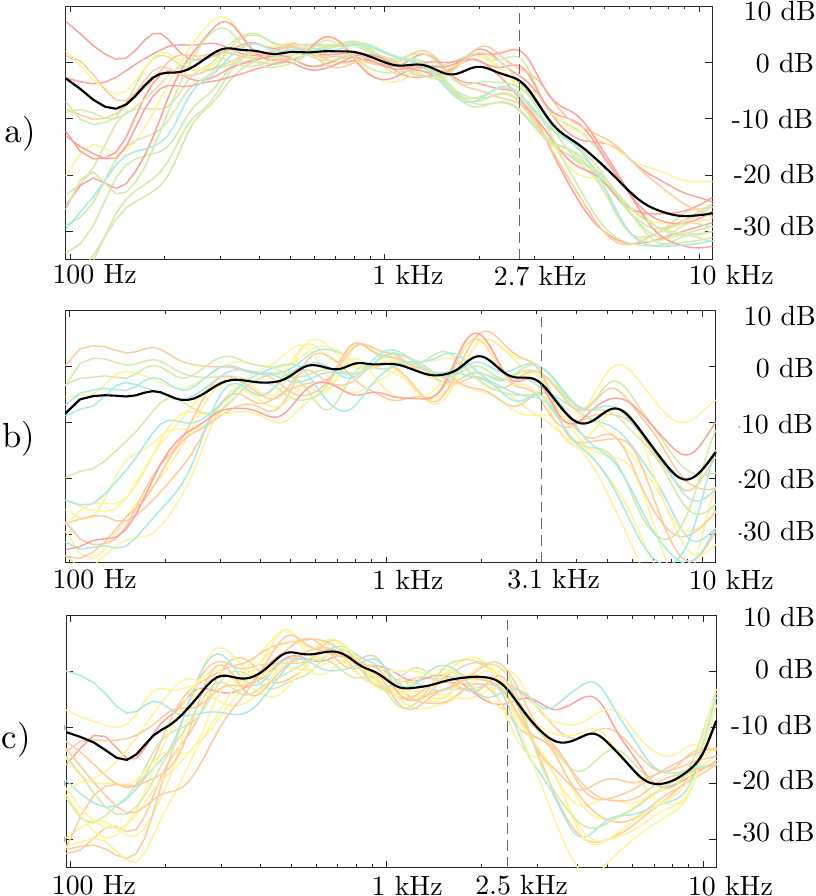}
    \caption{LTAS difference curves computed between six pre-1930s recordings and three contemporary recordings of (a) \textit{The Blue Danube Waltz}, (b) \textit{Carmen}, and (c) \textit{Humoresque}. Each colored line represents the difference between the LTAS of one of the six old recordings and one of the three modern ones, totalling 18 curves. The black line is the average of the difference curves, and the vertical dashed line marks its $-3$-dB point.}
    
    \label{ltas}
\end{figure}

To obtain prior knowledge to design our method, we analyzed the bandwidth of 78-RPM (rounds per minute) gramophone recordings, which we were interested in enhancing. To fully understand the frequency characteristics of these recordings, one must study the recording conditions of the time. However, due to the lack of international standards, the exact characteristics vary widely depending on the manufacturer, the publication date, the recording material, or possible equalization corrections made by recording engineers. Hence, the work of audio restoration is extremely hard, as restoration engineers now have to conduct a study on industrial archaeology for every single record they aim to restore \cite{copeland2008manual}. 

One of the main reasons for the limited bandwidth in old analog recordings are the disc-cutting lathes, used to record sound into the physical disc media \cite{copeland2008manual}. The most critical piece, the cutterhead, converts electric waveforms into modulations in a groove. The frequency response of the recording vary greatly depending on the cutterhead model, the speed of the record, or the shape of the stylus. The most commonly used cutterheads during the early 1920s produced a resonance frequency between 3\,kHz and 4\,kHz,  and above that the frequency response decayed rapidly. As a consequence, due to the poor signal-to-noise ratio (SNR) of the recordings, the high-frequency components above this resonance frequency were practically lost. With the introduction of Western Electric's electromechanical cutterhead in 1925, the frequency response could be considerably flattened, but the cutoff frequency could not be extended to above 5\,kHz \cite{copeland2008manual}. Over the years, better equipment was developed that allowed engineers to extend the recordable bandwidth of audio, thanks to many technological advances like motional feedback \cite{copeland2008manual}.

To analyze empirically the spectral characteristics of 78-RPM gramophone recordings, we conduct a study based on the LTAS. To do so, we collected six 78-RPM gramophone recordings of a given music piece, all of them containing a similar ensemble of instruments and dated from 1920 to 1930. For comparison, we also collected three contemporary broadband recordings of the same piece. The six old recordings were first denoised using our previously proposed method \cite{Moliner2022}. The LTAS was calculated for each of the recordings using the IoSR library \cite{iosr}, applying Gaussian smoothing per octave band. We then subtract the LTAS of the three contemporary recordings from each of the old versions to obtain a rough estimate of the frequency response of the recording. The resulting 18 difference LTAS curves are re-scaled so that their mean level between 500\,Hz and 2\,kHz is 0\,dB. 

Fig.~\ref{ltas} shows the computed difference LTAS curves and their average for three classical pieces: the orchestral piece \textit{The Blue Danube Waltz}, by Johann Strauss (Fig.~\ref{ltas}(a)); the opera piece \textit{L'amour est un oiseau rebelle}, from Carmen by Georges Bizet (Fig.~\ref{ltas}(b)); and \textit{Humoresque No. 7}, by Antonin Dvořák, played by string ensembles (Fig.~\ref{ltas}(c)). Although these plots do not give accurate information due to the averaging and the octave-band smoothing, they indicate a decaying trend starting at approximately 3\,kHz. The estimated $-3$-dB cutoff frequencies are 2.7, 3.1, and 2.5\,kHz for the above-mentioned historical recordings, as indicated in Fig.~\ref{ltas}.

 

\section{BEHM-GAN}
\label{sec:methods}


This section presents the BEHM-GAN, the proposed GAN-based method for the bandwidth extension of historical recordings. The generator model, the loss functions, and the three different discriminators used are first described, and their roles are also illustrated in Fig. 3. The dataset, the use of lowpass filters to simulate the loss of high frequencies, and the implementation of the training are also explained.

\begin{figure}
    \centering
    \includegraphics{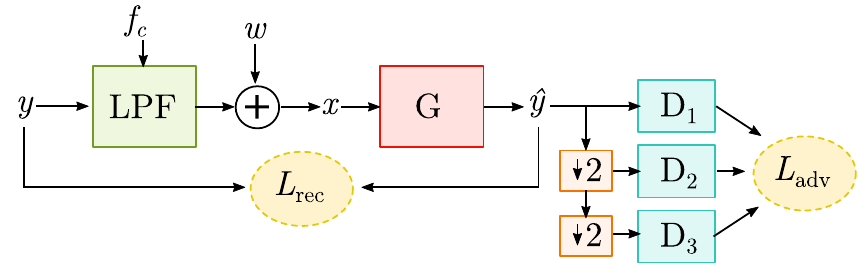}
    \caption{Proposed GAN-based training framework containing generator $G$, three time-domain discriminators $D_1$, $D_2$, and $D_3$, a variable lowpass filter (LPF), and additive noise $w$. The training is optimized with a composite of two losses: an adversarial loss $L_\text{adv}$ and a reconstruction loss $L_\text{rec}$. The down-sampling operators refer to strided average pooling with a kernel size of 4.}
    \label{training_framework}
\end{figure}

\subsection{Generator Model Architecture}
\label{sec:architecture}

\begin{figure*}
    \centering
    \includegraphics{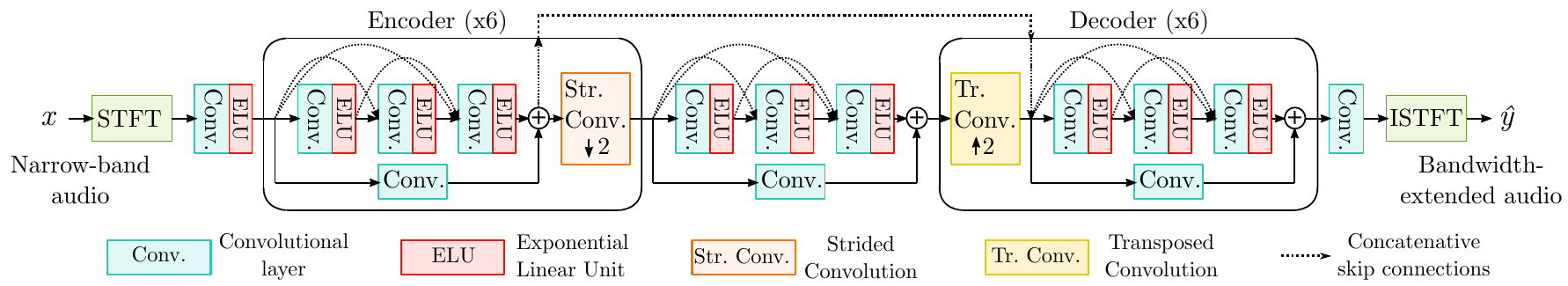}
    \caption{Proposed U-net-based architecture of the generator model, cf.~Fig.~\ref{training_framework}.}
    \label{generator}
\end{figure*}
We opt to work in the time-frequency domain to design our generator. Thus, the input bandlimited audio $x$ sampled at $f_\text{s}=22.05$\,kHz is first transformed by means of the short-time fourier transform (STFT). We use an FFT length of 1024 samples (46.11\,ms) with a Hamming window of the same size and a hop length of 256 samples (11.61\,ms). The resulting complex signal is converted to a real one by stacking its real and imaginary parts as separate channels. Then, assuming that the generator succeeds at maintaining the implicit phase information, the output signal can be directly converted back to the time domain using the inverse-STFT, without the need for a phase-recovery technique. We opted not to use a complex-aware neural network architecture \cite{trabelsi2018deep} as, in our experiments, it did not provide any clear benefit against the double-real representation. Moreover, complex-valued modules require twice the amount of computation, an issue that slowed down the training significantly. 

The generator architecture is based on the U-Net model \cite{Moliner2022} and is shown in Fig.~\ref{generator}. The architecture is formed by 2D-convolutions and Exponential Linear Unit non-linearities \cite{clevert_fast_2016} to capture time-frequency features from the spectrogram. We concatenate frequency-positional embeddings \cite{isik_poconet_2020} as an inductive bias to break the frequency-equivariance symmetry, which is implicit in 2D-convolutions. The architecture has an encoder-decoder structure with residual DenseNet blocks \cite{huang_densely_2018} as intermediate layers. The encoder coarsens the resolution at each layer using strided convolutions, sequentially increasing the number of channels. The decoder structure is symmetrical to the encoder and upsamples the resolution with transposed convolutions. The concatenative skip connections help to retain fine-grained details of the spectrogram. We refer the reader to the source code\footnote{https://github.com/eloimoliner/bwe\_historical\_recordings} for further details on the model implementation and the used hyperparameters.

\subsection{Training Objective}
\label{sec:training}

The generator is optimized with a composite of two losses, an adversarial loss $L_\text{adv}$ and an auxiliary reconstruction loss $L_\text{rec}$: 
\begin{equation}
   L_G=L_\text{adv} +\alpha L_\text{rec},
\end{equation}
where the coefficient $\alpha=0.4$ is a tuning hyperparameter used to combine the two loss terms. The value of $\alpha$ was optimized by grid search, using informal listening as the quality criterion. For the adversarial loss, we adopt the multi-scale discriminators $D_1$, $D_2$, and $D_3$ from MelGan \cite{kumar2019melgan}.
 
As indicated in Fig.~\ref{training_framework}, discriminator $D_1$ operates directly on the raw audio waveform, whereas the input waveforms of discriminators $D_2$ and $D_3$ are, respectively, downsampled by factors 2 and 4. Thus, each discriminator learns features in a different frequency range. Since the model operates at $f_\text{s}=22.05$\,kHz, $D_1$ observes frequency components up to the Nyquist limit $f_\text{s}/2=11.03$\;kHz, $D_2$ up to $f_\text{s}/4=5.51$\;kHz and $D_3$ only to $f_\text{s}/8=2.76$\;kHz. Although our main interest is to reconstruct the frequency components above 3.0\;kHz, using $D_3$ is still beneficial to stabilize the adversarial training and leads to better convergence. The architecture of each of the discriminators consists of a stack of grouped strided convolutions, and the down-sampling is performed by strided average pooling with a kernel size of 4, in the same way as in \cite{kumar2019melgan}. Time-domain discriminators are highly sensitive to phase mismatches in the data. This is a very convenient property when using a spectrogram-based generator, since maintaining the phase coherence when the audio data is transformed to the complex STFT domain may be problematic. 

In this work, we apply the least-squares GAN objective \cite{mao2017least}. The adversarial loss for the generator is then defined as
\begin{equation}
L_{\text{adv}}= \mathbb{E}_{\hat{y_k}}
\left[
\sum_k (D_k(\hat{y_k})-1)^2
\right],
\end{equation}
where $\mathbb{E}$ is the expectation operator. The discriminators are optimized by minimizing the loss function:
\begin{equation}
L_{D_k}= \frac{1}{2}\mathbb{E}_{y_k}\left[(D_k(y_k)-1)^2\right]+\frac{1}{2}\mathbb{E}_{{\hat{y_k}}}\left[D_k(\hat{y_k})^2\right].
\end{equation}

We use the multi-resolution STFT loss \cite{yamamoto2020parallel} as the auxiliary reconstruction loss $L_\text{rec}$. This loss is defined as the expectation of the sum of two terms, $L^{(m)}_{\text{sc}}$ and $L^{(m)}_\text{mag}$, at $M$ different frequency resolutions as
\begin{equation}
L^{(m)}_{\text{rec}}= \mathbb{E}_{y,\hat{y}}
\left[
\frac{1}{M}
\sum_{m=1}^M
L^{(m)}_{\text{sc}} +L^{(m)}_{\text{mag}}
\right].
\end{equation}

\noindent The spectral convergence term $L^{(m)}_{\text{sc}}$ and the log magnitude distance term $L^{(m)}_\text{mag}$ are defined, respectively, as:
\begin{equation}
    L^{(m)}_{\text{sc}}=\frac{\lVert \;|Y^{(m)}|  -|\hat{Y}^{(m)}|
    \;\rVert_\text{F}}{\lVert \;|Y^{(m)}|\; \rVert_{\text{F}}}
\end{equation}
and
\begin{equation}
    L^{(m)}_\text{mag}=\frac{1}{S}
    \lVert
    \log |Y^{(m)}|
    -\log |\hat{Y}^{(m)}|
    \rVert_{1},
\end{equation}
where $Y^{(m)}$ and $\hat{Y}^{(m)}$ are the STFTs of the signals $y$ and $\hat{y}$, respectively, using an analysis window of length $m\in{\{256,512,1024,2048\}}$,$\lVert \cdot \rVert_\text{F}$ is the Frobenius norm, $\lVert \cdot \rVert_1$ is the L1 norm, and $S$ is the total number of STFT bins. Note that this reconstruction loss is only aware of the magnitude differences in the spectrogram, as we rely on the adversarial loss term to deal with the phase information.




\subsection{Dataset}
\label{sec:dataset}

We train and evaluate our method using solo piano classical music. Doing so, we reduce the difficulty of the problem by limiting the variance in the training data. Piano sounds are a convenient choice for evaluating bandwidth-extension algorithms, since they contain both transient and tonal components. Moreover, since the piano is one of the most common musical instruments for solo performances, a large quantity of contemporary and historical solo piano recordings is publicly available. Considering that classical music is a genre that has practically remained unchanged over time, we avoid introducing a major divergence between the training and target distributions.

We collected our training data from the solo piano pieces of the MusicNet dataset \cite{thickstun_learning_2017}, but discarded some of the older recordings as they contained heavy background noise and the audio quality was suboptimal. The training set contains 14.4\,h of broadband piano classical music. A separate test set with 1.1\,h of broadband piano music is used for the objective and subjective evaluation metrics that require a reference signal (Sec. \ref{sec:obj} and Sec. \ref{sec:subj_synth}). The music pieces included in the test set are not present in the training set.

A test set of real historical recordings was also collected to compute the objective and subjective evaluation metrics that do not require a reference signal (Sec. \ref{sec:obj} and Sec. \ref{sec:subj_old}). It consists of six historical solo piano recordings extracted from ``The Great 78 Project" \cite{78project}, a large collection of publicly available digitized 78-RPM gramophone records \cite{archive}.

\subsection{Lowpass Filter Generalization}
\label{sec:filtreg}


Since the frequency responses in historical music recordings are far from being deterministic, we apply a lowpass filter with a randomized cutoff frequency $f_\text{c}$ to the training data. We parameterize the cutoff frequency with a normal distribution, whose mean and standard deviation are set up empirically to be a rough estimate of the frequency responses in the gramophone recordings in the 1920s.

\begin{figure}
    \centering
    \includegraphics{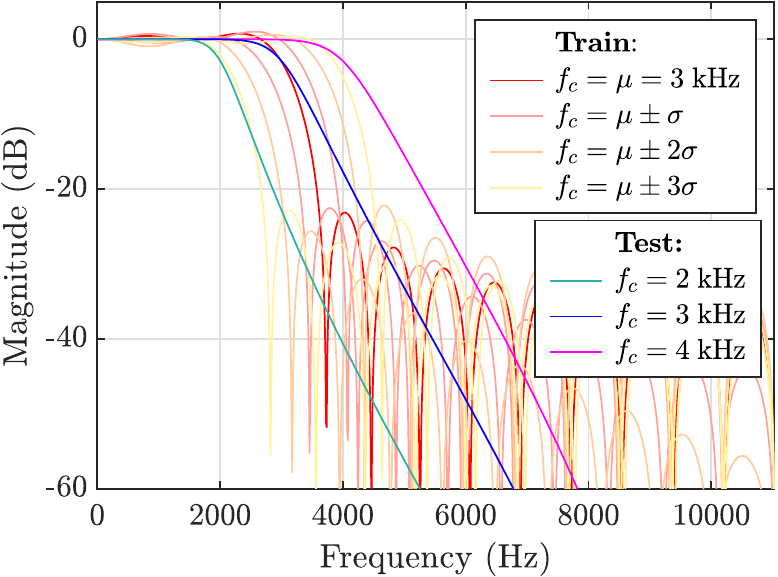}
    \caption{Magnitude responses of lowpass filters used for training and testing. The training filters are FIR with a randomized cutoff frequency, while the testing filters are IIR with a fixed cutoff.}
    \label{all_filters}
\end{figure}

 Based on our findings from the spectral analysis in Sec.~\ref{sec:Spectral_Analysis}, we set the mean cutoff frequency to $\mu_{f_\text{c}}=3.0\text{\,kHz}$ and the standard deviation to $\sigma_{f_\text{c}}=300\text{\,Hz}$. The value of $\sigma_{f_\text{c}}$ is the result of a trade-off off bias against variance in the model. In other words, a model trained with a larger $\sigma_{f_\text{c}}$ would probably generalize to a wider range of cutoff frequencies. However, the resulting quality would likely diminish due to the increase in variance in the data, making the optimization more challenging and unstable. The lowpass filters are 25th-order FIR filters using the windowing method with the Kaiser window ($\beta=1$). The magnitude responses of the FIR filters used for training are presented in Fig.~\ref{all_filters}. FIR filters have the convenient advantage that they can be efficiently implemented by applying convolution, thus not demanding much extra computation during training.

The idea of applying variable-band filters was also studied by Li \emph{et al}. \cite{li_real-time_2021}, with the difference that they used a uniform distribution instead of a normal one. Randomizing the cutoff frequency of the filter indeed helps to increase the robustness of the model in different frequency ranges, but, as we show with our experiments in Sec.~\ref{sec:subj_old}, randomization is certainly not enough if we want to successfully make inferences in historical recordings. 

\begin{figure}
    \centering
    \includegraphics{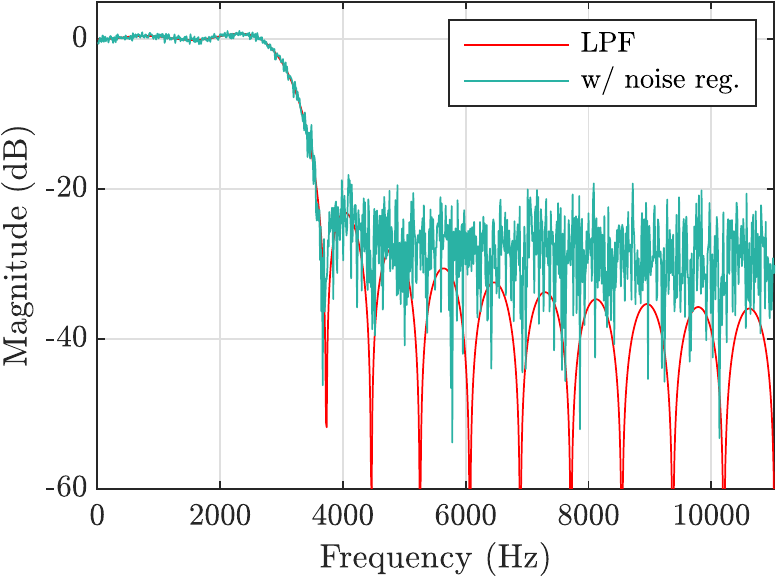}
    \caption{Magnitude response of one of the lowpass filters used during training and its equivalent response after applying the noise regularization.}
    \label{noisefilter}
\end{figure}

To further regularize the model, we apply a well known regularization approach \cite{bishop1995training} and corrupt the lowpass-filtered training data by adding a small amount of Gaussian white noise $w$ having zero mean and a fixed power $\sigma^2=-30\,\text{dBFS}$ (decibels relative to full scale)  directly to the raw audio signal. The added noise diffuses the magnitude response of the filter, as Fig.~\ref{noisefilter} demonstrates, and successfully enforces the model to generate new high-frequency content instead of overfitting the filter shape. The use of the additive noise during training also encourages the generator to focus only on the most prominent musical features making the model robust to the minuscule denoising residuals that it may encounter while making inferences in historical recordings. Furthermore, this noise regularization strategy injects stochasticity into the model. Given that we do not add a latent vector $z$ to the generator, this extra noise allows the generator to produce stochastic outputs. The fixed value of $\sigma^2=-30\,\text{dBFS}$ is chosen in consistence with the training lowpass filters, in such a way that the noise level is sufficiently high to mask the remaining information in the side lobes of the filter (see Fig.~\ref{noisefilter}). We observed that using higher values of $\sigma^2$ was often detrimental to the resulting audio quality, as some unwanted noisy residuals were still present in the output.

\subsection{Making Inferences in Historical Recordings}
\label{sec:inference}

Using the proposed regularization, the generator can be applied to make inferences on out-of-distribution historical music recordings. Our inference pipeline builds on previous work on denoising \cite{Moliner2022}, as illustrated in Fig.~\ref{inference}. The original noisy recordings are first denoised to suppress clicks, hisses, and other additive disturbances. Then, the denoised recordings are bandwidth-extended by directly applying the pre-trained STFT-based generator.

In the same way as during training, the noise regularization could be added at the inference stage before feeding the denoised recording to the bandwidth-extension generator. As discussed in Sec.~\ref{sec:obj}, this step helps to achieve better objective metrics. However, in the majority of the tested cases, no perceptual differences were noticed with or without noise during inference and hence it is left as an optional step.


\subsection{Implementation Details}
\label{sec:details}

The used sampling frequency $f_\text{s}=22.050$\;kHz sets the upper limit of processing to about 11\,kHz. This choice makes the training fast and still leaves a wide range from about 3 to 11\,kHz for bandwidth extension. For training, we used batches of four audio segments, each with a duration of 5\,s. Nevertheless, due to the nature of convolutional neural networks, the input length can be set arbitrarily during inference. With the goal to make the model robust to different volume (loudness) levels, we also apply a uniformly random gain, set between $-6$\,dB and $4$\,dB for each input signal. We did not find using batch normalization or weight normalization beneficial to the generator. The discriminators, however, are weight-normalized \cite{salimans2016weight}.

We use the Adam optimizer \cite{kingma_adam_2017} with the parameters $\beta_1=0.5$ and $\beta_2=0.9$ to train both the generator $G$ and the discriminators $D_k$. The training is divided into two separate stages. First, we train $G$ for 10,000 steps with a learning rate of $1\times10^{-4}$ using only the reconstruction loss $L_{\text{rec}}$. This step guarantees that the model learns to apply an identity mapping to the low-frequency components before including the adversarial discriminators into the training loop. Then, we decrease the learning rate to $1\times10^{-5}$, incorporate the adversarial loss $L_{\text{adv}}$, and continue training for 300,000 steps. During the second stage, the discriminators $D_k$ are updated twice for every step taken by the generator, using a learning rate of $1\times10^{-4}$. The training took, on average, two days to complete on a single Tesla V100 GPU in Triton, Aalto University's computing cluster.

\section{Experiments and Results}
\label{sec:experiments}

This section evaluates the quality of the bandwidth-extension using both objective and subjective experiments.

\subsection{Comparison Models}

\begin{figure*}[t]
\centering
    \includegraphics[scale=0.15]{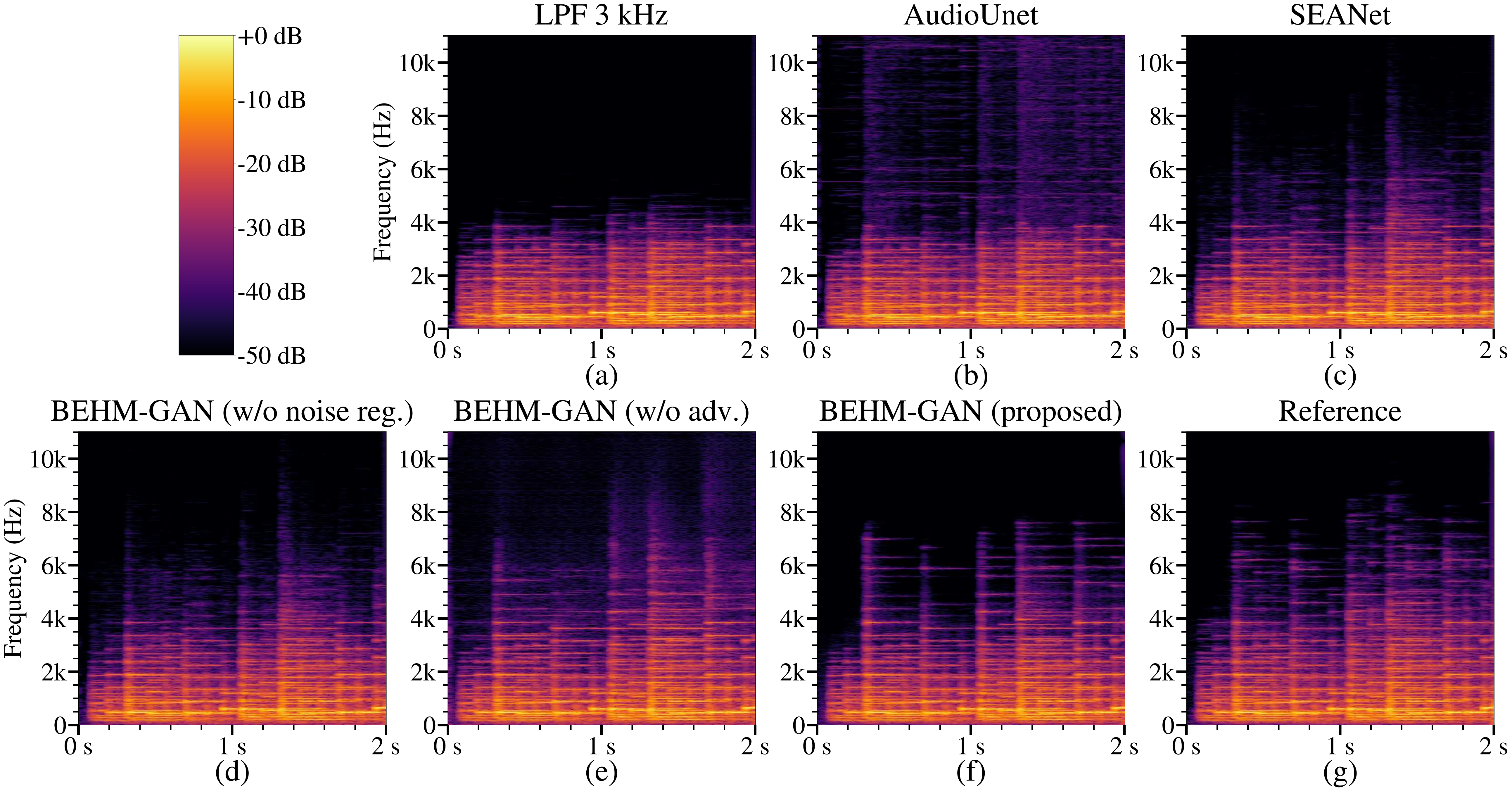}
    \caption{Spectrograms of (a) a lowpass filtered reference signal, (b), (c), (d), (e), (f) its bandwidth-extended versions, and (g) a reference modern piano recording. }
    \label{sim_spectrograms}
\end{figure*}
\begin{figure*}[t]
\centering
    \includegraphics[scale=0.15]{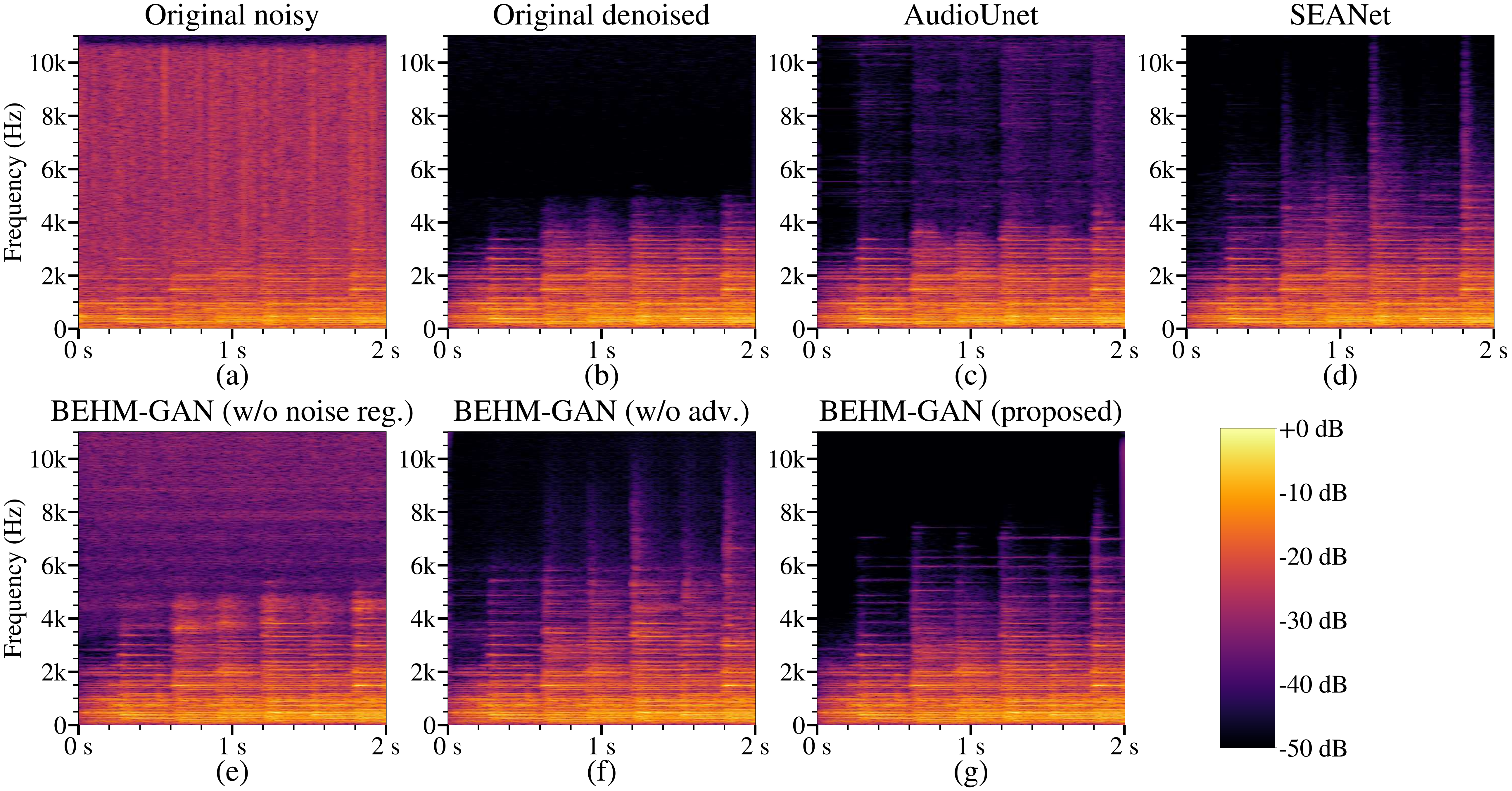}
    \caption{Spectrograms of (a) an original noisy historical recording, (b) its a denoised version, and (c), (d), (e), (f), (g) the bandwidth-extended versions of (b). }
    \label{real_spectrograms}
\end{figure*}

We compare our proposed method with two baseline models, AudioUnet \cite{kuleshov2017audio} and SEANet \cite{li_real-time_2021}. Considering that these baselines were not designed to do bandwidth extension in historical recordings, their training had to be adapted to perform well on this task. Unless otherwise specified, the compared baselines were trained using the same methodology as the BEHM-GAN (Sec.~\ref{sec:methods}), using the same lowpass filters and noise regularization.

AudioUnet is a supervised model based on a time-domain U-Net \cite{kuleshov2017audio}, which was originally presented to enhance both speech and piano music signals sampled at 16\,kHz. The model is optimized using a reconstruction L2 loss between the resulting output waveform and the original unfiltered audio. We use the PyTorch implementation of the model released by Sulun and Davies\footnote{https://github.com/serkansulun/deep-music-enhancer}.

SEANet is a GAN-based model that was used for speech enhancement \cite{tagliasacchi2020seanet} and speech bandwidth extension \cite{li_real-time_2021}. Its generator is a time-domain U-Net with dilated convolutions at the intermediate layers. An effort was made to replicate the implementation details from the larger non-real-time model evaluated in \cite{li_real-time_2021}. Given that the original SEANet utilizes very similar, if not the same, time-domain discriminators as in this paper, we opted to train SEANet using the same training objective as ours. This gave us better performance than the ``feature'' loss the authors originally applied and allowed us to directly evaluate the effects of using a spectrogram-based generator versus a time-domain one.

We also experimented with TFiLM \cite{birnbaum2021temporal}, the GAN-based HiFi-GAN \cite{su_bandwidth_2021}, and the diffusion probabilistic NU-Wave models \cite{lee_nu-wave_2021}. However, we did not obtain positive results using these methods for bandwidth-extending historical recordings. We hypothesize that, in contrast to speech, the aforementioned models may not be well suited for processing music instead of speech or that further hyperparameter optimization is necessary to adapt the models to our training methodology. We decided not to include these baselines in the formal evaluation to avoid reporting misleading results and to not overload the number of listening conditions in the subjective evaluation.

So as to understand the effects of the main components of our approach, we also include four ablated versions of our model in the formal evaluation. Firstly, we study the importance of noise regularization by training a model without adding white noise (w/o noise reg.). Secondly, we switch the adversarial training objective for an L2 reconstruction loss in the complex-spectrogram domain (w/o adv.). The multi-resolution STFT loss $L_{\text{rec}}$ is not used, being ineffective if used alone, since the phase information in the spectrogram is ignored. We also report, although only with objective metrics, the effect of adding noise to the input audio signal during inference, in the same way as we do during training (with noise inf.). Finally, we also included in the objective evaluation a model based on a complex-aware architecture \cite{trabelsi2018deep}, which consists of the same architecture from Fig.~\ref{generator}, but replacing each of the convolutional blocks with their complex-valued counterparts.

Figs.~\ref{sim_spectrograms} and \ref{real_spectrograms} show a visual comparison of the spectrogram representations of the compared models in a modern-lowpass filtered example and an old recording, respectively. Figs.~\ref{sim_spectrograms}(a) and \ref{real_spectrograms}(b) present, respectively, the bandlimited input signals for the examples, and the other spectrograms visualize how each method recreates the missing high-frequency content. The reference signal is added in Fig.~\ref{sim_spectrograms}(g) so that the results can be compared with the original real spectrogram. As is evident, some of the compared methods produce a more realistic spectra than others. A reference signal is unavailable for the old recording in Fig.~\ref{real_spectrograms}. In this case, we additionally present the spectrogram of the noisy old recording before applying the denoiser in Fig.~\ref{real_spectrograms}(a), which reveals that the old recordings contains only noise and distortion above about 4\,kHz.

Fig.~\ref{sim_spectrograms}(f) and Fig.~\ref{real_spectrograms}(g) show the spectrogram of the results of the proposed method. By comparing them with the input bandlimited signals (Fig.~\ref{sim_spectrograms}(a) and Fig.~\ref{real_spectrograms}(b), respectively), one can observe how the bandwidth has been practically doubled. Nevertheless, almost no new frequency has been generated above 8\,kHz. Another issue that can be perceived is the fact that the model is sometimes too aggressive in building up high frequencies, meaning that sometimes they may start too early. We hypothesize that the problem may come from the time-frequency processing, as the FFT window may be smearing the transients of the piano notes.

\subsection{Objective Evaluation}
\label{sec:obj}

\begin{table*}[t]
\centering
\caption{Objective Metrics, Where Lower is Better for all Cases. The Best Result in Each Column is Higlighted}
\label{objectivemetrics}
\begin{tabular}{@{}l|lll|lll|lll|c@{}}
\toprule
                       & \multicolumn{3}{c|}{$f_\text{c}=2$\;kHz}                 & \multicolumn{3}{c|}{$f_\text{c}=3$\;kHz}                 & \multicolumn{3}{c|}{$f_\text{c}=4$\;kHz}                 & \multicolumn{1}{c}{Historical recording} \\ 
                       & \textbf{LSD}  & \textbf{VGGish} & \textbf{FAD}  & \textbf{LSD}  & \textbf{VGGish} & \textbf{FAD}  & \textbf{LSD}  & \textbf{VGGish} & \textbf{FAD}  & \textbf{FAD}                   \\ \midrule
LPF/Original           & 1.01          & 5.38            & 4.96          & 0.82          & 3.93            & 3.48          & 0.71          & 3.43            & 2.07          & 2.16                           \\ 
AudioUnet              & 0.99          & 4.57            & 3.51          & 0.84          & 3.99            & 1.96          & 0.79          & 3.73            & 1.56          & 2.88                           \\
SEANet                 & 0.99          & 4.41            & 3.87          & 0.81          & 3.38            & 1.18          & 0.75          & 3.21            & 1.05          & 1.66                           \\
BEHM-GAN (proposed)               & \text{0.87} & \text{4.15}   & \text{3.44} & \textbf{0.71} & \text{3.21}   & \text{0.70} & \textbf{0.66} & \text{3.01}   & \text{0.70} & \text{1.26}                  \\
\;\;\;\;w/o noise reg. & 0.92          & 4.45            & 5.14          & 0.73          & 3.54            & 2.60          & 0.7           & 3.47            & 1.04          & 6.00                           \\
\;\;\;\;w/o adv.       & 1.00          & 5.42            & 5.18          & 0.86          & 3.99            & 2.95          & 0.75          & 4.02            & 1.91          & 2.23                           \\ 
 \;\;\;\;with complex conv.              & \textbf{0.85} & \text{4.21}   & \textbf{1.02} & \textbf{0.71} & \text{3.26}   & \textbf{0.53} & \textbf{0.66} & \text{2.93}   & \textbf{0.47} & \text{1.67}                  \\
\;\;\;\;with noise inference & \textbf{0.85}    & \textbf{3.98}   & \text{3.34} & \textbf{0.71}     & \textbf{3.00}   & \text{0.68} & \textbf{0.66}      & \textbf{2.76}   & \text{0.64} & \textbf{1.12}                                \\ \midrule
Reference              & -             & -               & 0.58          & -             & -               & 0.58          & -             & -               & 0.58          & -                              \\ \bottomrule
\end{tabular}
\end{table*}
To conduct the objective evaluation, we utilize the test set described in Sec. \ref{sec:dataset}, which comprises 70 min. of modern broadband classical music recordings. All the audio signals from the test set were resampled at the rate $f_\text{s}=22.05$\,kHz and were split into non-overlapping 5-s segments. A sixth-order Butterworth lowpass filter is applied at the fixed cutoff frequency of 2\,kHz, 3\,kHz, and 4\,kHz to imitate different bandwidth limitations. Note that the testing filters are purposely different from the training filters in order to evaluate the models in out-of-distribution filtering conditions. The magnitude responses of the three Butterworth filters used for testing are shown in Fig.~\ref{all_filters}, together with the filters used during training. 
  
The proposed method and the aforementioned baselines are evaluated using three objective metrics: log-spectral distance (LSD), VGG distance (VGG), and Fréchet Audio Distance (FAD) \cite{kilgour2019frechet}.

\subsubsection{Log-Spectral Distance}
The LSD, a frequency-domain metric that has been popularly used in bandwidth-extension literature \cite{kuleshov2017audio,Wang2021,su_bandwidth_2021,Eskimez2019Adversarial}, is defined as: 
\begin{equation}
   \text{LSD}=\frac{1}{T} \sum_{t=1}^T \sqrt{
   \frac{1}{K}
   \sum_{k=1}^K
   \left(
   \log |Y_{t,k}|^2- \log |\hat{Y}_{t,k}|^2
   \right)^2
   },
\end{equation}
where $Y_{t,k}=\text{STFT} (y)$ and $\hat{Y}_{t,k}=\text{STFT} (\hat{y})$  are the STFTs of the reference $y$ and the bandwidth-extended audio signal $\hat{y}$, respectively. For the STFT computation, an analysis window of $K=2048$ samples and a hop length of 512 samples is used. 

\subsubsection{VGG Distance}
This metric is defined as the L2 distance between pairs of individual embeddings given by the VGGish network \cite{hershey2017cnn}.
The VGGish network has been pre-trained for large-scale audio classification. Thus, this metric is expected to provide a distance measure focusing on the higher level features of the audio data. This metric was previously used to evaluate music denoising \cite{li2020learning} and bandwidth-extension \cite{sulun_filter_2020} methods.
 
\subsubsection{Fréchet Audio Distance}
FAD \cite{kilgour2019frechet} has been adapted for audio from the Fréchet Inception Distance (FID),  a frequently used metric to evaluate image-generative models. This metric also uses the VGGish embeddings to compare the statistics between two collections of audio. FAD fits a multivariate normal distribution to a collection of background $\mathcal{N}(\mu_\text{b}, \Sigma_\text{b})$ and  evaluation $\mathcal{N}(\mu_\text{e}, \Sigma_\text{e})$ embeddings. Then, the Fréchet distance between both distributions is defined as: 
\begin{equation}
\text{FAD} = \lVert  \mu_\text{b} -  \mu_\text{e} \rVert_2 + \text{tr}(\Sigma_\text{b} + \Sigma_\text{e} - 2\sqrt{\Sigma_\text{b} \Sigma_\text{e}}),
\end{equation}
where $\lVert \cdot \rVert_2$ is the L2 norm and tr($\cdot$) is the trace of a matrix. Being reference-free, we avoid computing FAD on paired data by dividing the test set into two equally-sized splits. One of them is used to compute the background statistics, whereas the second is used to evaluate the different models. 

 As FAD is reference-free, it is also used to evaluate the bandwidth-extension performance in real historical recordings. Similarly, the FAD is computed from 15 min of historical recordings using the same background statistics. The six historical recordings that we use for evaluation were extracted from ``The Great 78 Project" \cite{78project}, as mentioned in Sec. \ref{sec:dataset}. However, since the FAD has the limitation of working at the sampling frequency of 16\;kHz, which differs from that at which the BEHM-GAN operates (22.05\;kHz), the audio signals must be resampled before computing this metric. As a consequence, the FAD only observes frequency components up to the Nyquist limit of 8\;kHz, missing some high-frequency details. The study of the design of broadband reference-free audio quality metrics is left as future work.

No SNR-related metric is used for the objective evaluation because they are extremely sensitive to phase misalignments between pairs of data. Since the training and testing filters have a different phase response, the waveform representations of the reference and the bandwidth-extended audio can differ greatly, although they may sound similar. For this reason, the SNR results do not correlate with perceptual audio quality and are discarded.

\begin{table}[]
\caption{Comparison Study of Adding Noise at the Inference Stage. The Best Result in Each Row is Highlighted}
\centering
\label{noiseFAD}
\begin{tabular}{@{}llll@{}}
\toprule
                                                                                       &           & \multicolumn{2}{c}{FAD}              \\
                                                                                       &           & w/o noise at inf. & with noise at inf. \\ \midrule
\multirow{3}{*}{\begin{tabular}[c]{@{}l@{}}Lowpass filtered\\ $f_\text{c}=3$\,kHz\end{tabular}} & AudioUnet & \textbf{1.96}     & 3.64             \\
                                                                                       & SEANet    & 1.18              & \textbf{1.09}    \\
                                                                                       & BEHM-GAN  & 0.70              & \textbf{0.68}    \\ \midrule
\multirow{3}{*}{\begin{tabular}[c]{@{}l@{}}Historical\\ recordings\end{tabular}}       & AudioUnet & \textbf{2.88}     & 6.74             \\
                                                                                       & SEANet    & \textbf{1.66}     & 1.91             \\
                                                                                       & BEHM-GAN  & 1.26              & \textbf{1.12}    \\ \bottomrule
\end{tabular}
\end{table}
The objective results are tabulated in Table~\ref{objectivemetrics}. Note that a reference condition for the FAD metrics is added in the lowpass filtered results. This refers to computing the FAD between the two test splits of broadband piano recordings. Thus, the reference results can be considered as a lower bound of the FAD metric.

The proposed method outperforms the two compared baselines in all the evaluated conditions and, most importantly, improves over the "LPF/Original" condition, where no bandwidth extension was applied. The performance decreases when the noise regularization is not used, and also when the model is trained without the adversarial losses, proving that these are critical features of the model within the context of the results obtained. We will return to this point when analyzing the subjective evaluation results in Sec.~\ref{sec:subj_old}. 

 The BEHM-GAN obtains the most substantial improvement when $f_\text{c}=3\,$kHz, which corresponds to the mean cutoff frequency used for the training filters. The model also generalizes well when the cutoff frequency is higher ($f_\text{c}=4\,$kHz). However, when decreasing the cutoff frequency ($f_\text{c}=2\,$kHz), the metrics deteriorate a little as the task gets considerably harder. Nevertheless, even in this case, a significant improvement is seen in all three metrics with respect to the unprocessed lowpass filtered condition. The objective metrics also show that the FAD is decreased by a factor of two when the BEHM-GAN is evaluated with real historical recordings, implying that the model is able to generalize in this real-world case.

Substituting the convolutional layers for complex convolutions \cite{trabelsi2018deep} does not produce an improvement in the reference-based objective metrics (LSD and VGGish). This configuration improves the FAD score, but only for the synthetic filtered data and not for the real historical recordings. In informal listening, we did not find any perceptual improvement in the results. As a consequence, we opt to use the double-real representation instead, given that the complex-valued layers require twice the amount of computation.

The metrics of the proposed method consistently improve when noise regularization is added at the inference stage. This is an expected result given that, in this case, the test example gets closer to the training distribution. However, as seen in Table~\ref{noiseFAD}, this does not necessarily happen with the other baseline methods. We noticed that AudioUnet and, to a lesser extent, SEANet sometimes do not succeed at completely suppressing the extra added Gaussian noise. No perceptual improvement was observed in the BEHM-GAN with or without noise added during inference or not. As a consequence, to make a fair comparison, only versions with no extra noise are included in the subjective evaluations.

\subsection{Subjective Evaluation of Synthetic Filtered Data}\label{sec:subj_synth}

\begin{figure*}[t]
\includegraphics[]{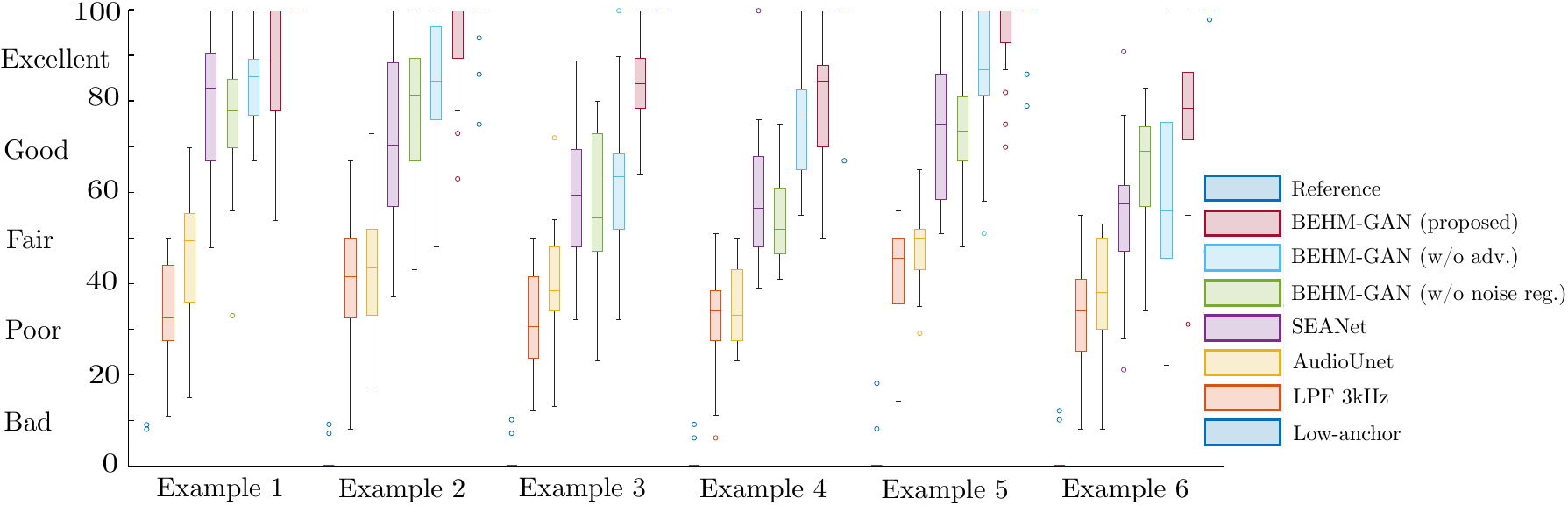}
\caption{Box-plot visualizations of the listening test results on synthetic lowpass filtered recordings.}
\label{simplots}
\end{figure*}

Since finding an objective metric that correlates with human perception is not easy, a formal subjective evaluation is conducted. We designed a blind listening test structured as two consecutive sessions, one to assess the performance of our method with simulated lowpass filtered piano music and the other to evaluate the performance in real historical piano recordings, which is detailed in Sec.~\ref{sec:subj_old}.

The first part was designed following the MUSHRA recommendation \cite{ITURmushra} with the purpose to evaluate the bandwidth-extension performance in synthetic lowpass filtered recordings. The listeners had to grade, on a perceptual scale from 0 to 100, the audio quality in eight different conditions. The reference signal was a contemporary broadband recording, expected to be rated as 100. Another condition was a lowpass filtered version of the reference at the cutoff frequency of 3\,kHz (LPF 3kHz), applying the same Butterworth filter that was used in the objective evaluation. The rest of the conditions were five different bandwidth-extended versions of the lowpass filtered signal, using the same methods as in Sec.~\ref{sec:obj}. Also included as a low anchor is an easy-to-recognize poor-quality signal lowpass filtered at 1.5\,kHz, which was expected to be graded as 0 by the listeners. We included six different 10-s piano music examples, repeated twice in random order, forming a total of 12 pages in the MUSHRA test. The audio examples included in the test are available listen at the companion webpage\footnote{http://research.spa.aalto.fi/publications/papers/ieee-taslp-behm-gan/}.

Altogether, 13 listeners participated in the listening test. However, one participant was discarded from the first part because they did not identify the reference in more than 15\% of the occasions, as recommended in \cite{ITURmushra}. All subjects had previous experience in formal listening tests, three of them were female, and their average age was 29 years. None of the participants reported of known hearing defects. The two test sessions took, on average, 45 min to complete. The experiment was conducted in the sound-proof listening booths of the Aalto Acoustics Lab, providing the same isolated listening conditions for all subjects. The listening test was implemented using the webMUSHRA interface \cite{schoeffler2018webmushra} that allowed the listeners to set loops if they wanted to focus on particular short passages of the audio signal. This feature was particularly useful for some participants, since the most noticeable differences between the conditions were localized in certain details, such as at the attack transients of the piano tones.


The results of the first session are plotted in Fig.~\ref{simplots}. Table~\ref{diffs_sim} presents the distances between the median scores of the BEHM-GAN and the compared conditions. The level of statistical significance given by a paired t-test is also marked. Examples 3, 4, and 6 contained intense \textit{fortissimo} piano passages, where the effect of the bandwidth limitation was easily audible. This explains why all the compared models obtained consistently lower ratings in these examples than in Examples 1, 2, and 5, which were softer and contained less high-frequency content. 

For all six examples, the proposed method obtained higher median scores than the other evaluated conditions. As indicated in paired t-tests (see Table~\ref{diffs_sim}), most differences are statistically significant. The subjects easily identified the reference in all the examples except examples 2 and 5, where a large proportion of listeners rated the BEHM-GAN with the maximum score of 100.

The worst-rated model was AudioUnet, which introduced some annoying aliasing artifacts that can be seen at the upper part of Fig.~\ref{sim_spectrograms}(b). SEANet worked significantly better but still produced a slightly distorted sound. The ablated versions of the proposed method obtained relatively good scores but, in the majority of the cases, were outperformed by the full BEHM-GAN model.

\subsection{Subjective Evaluation of Historical Recordings}
\label{sec:subj_old}
\begin{figure*}[t!]
\includegraphics[]{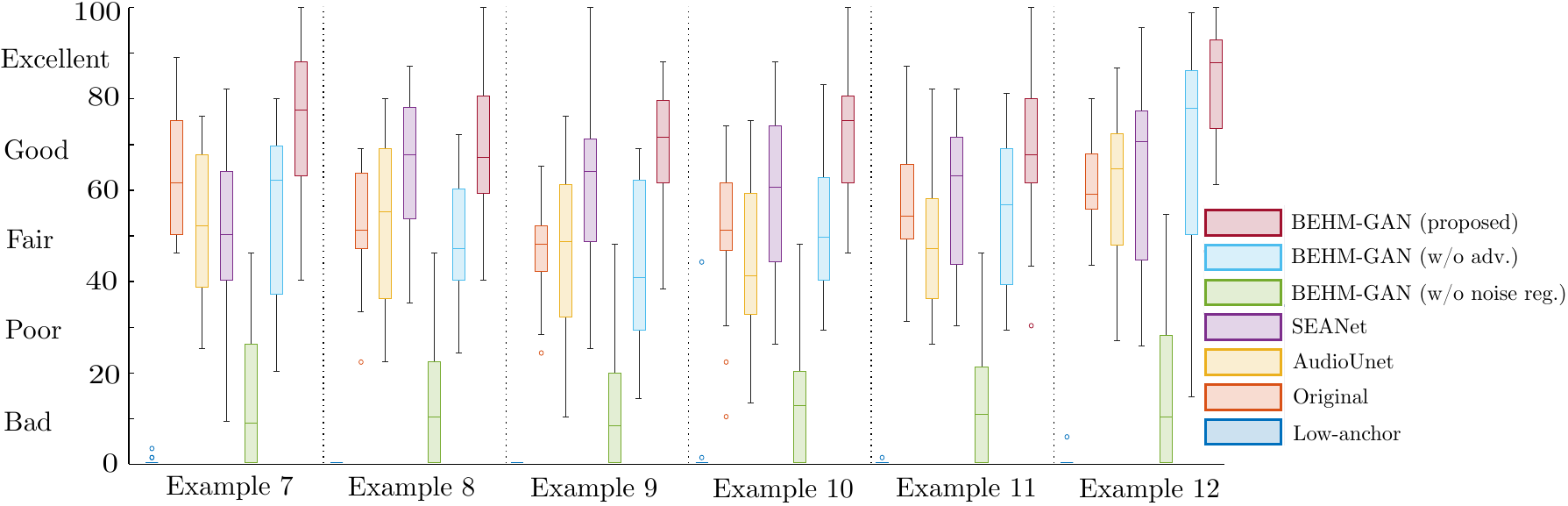}
\caption{Box-plot visualizations of the listening test results on real historical recordings.}
 \label{realplots}
\end{figure*}
The goal of the second session of the listening test was to evaluate the performance of the compared models in real historical recordings. The test method was a modified version of MUSHRA, where the audio examples were historical piano recordings. Since a broadband reference is unavailable, the reference presented to the listeners was, as in \cite{damskagg2017audio}, the unprocessed signal, in our case a denoised bandlimited gramophone recording. The tested conditions were the various bandwidth-extended versions of the reference, using the same methods as in the previous session. The same low anchor was included with the expectation of it being graded as 0, forming a total of seven conditions. The participants were asked to grade the audio quality for each of the conditions on a scale from 0 to 100 with the same criteria as in the first session, where 100 corresponds to a hypothetical perfect version of the reference. Thus, the participants were discouraged to rate any of the conditions with a score of 100 unless the quality was considered enhanced in a perfectly realistic manner. The second session included six 10-s examples of historical piano recordings, also repeated twice, and the total number of pages was 12.

 The experiment was conducted consecutively after the first session in the same conditions, with a short break between the two sessions. The test participants were also the same, except one who had to be discarded from the second part as they misunderstood the test question. This session took, on average, 20 min to complete.

The results of the second session are presented in Fig.~\ref{realplots} and Table~\ref{diffs_real}. Given that a higher ``excellent'' anchor was not available, the resulting scores contain more variance. Despite the wider confidence intervals, valuable conclusions can be extracted on how well each model generalizes to real historical recordings. 

The BEHM-GAN obtained significantly better results than the original recording in all the examples, implying that the bandwidth extension improved the sound quality. The other compared models introduced some distortion artifacts that the listeners sometimes evaluated negatively. The proposed method obtained marginally better scores than the rest of the conditions in all the examples except example 8, where SEANet received a slightly higher median score. Nevertheless, the BEHM-GAN outperformed SEANet in four out of six examples, as is evident from Table~\ref{diffs_real}. After finishing the test, some participants commented that the enhancement was often more noticeable with the time-domain SEANet model. However, the proposed method produced more realistic results, despite being more conservative.

Both ablated conditions suffered from a decline in performance. In particular, when noise regularization was ablated (BEHM-GAN w/o noise reg.), the resulting metrics were disappointing. These results can be explained by looking at the example in Fig.~\ref{real_spectrograms}(e), which looks noisy and distorted. We hypothesize that the model learned to invert the frequency response of a hypothetical lowpass filter but it failed when it encountered an out-of-distribution example where no such filter was applied. By closely inspecting the spectrogram of Fig.~\ref{real_spectrograms}(e), the model is observed to significantly boost the frequency bands above 4\,kHz and introduce some horizontally-shaped noisy components that resemble the side lobes of the lowpass filters seen during training. These results show that noise regularization is critical for the good performance of our system. Another observation is that the non-adversarial condition (BEHM-GAN w/o adv.) obtained worse scores in this case, implying that the proposed adversarial training objective is highly beneficial to generate a more realistic enhancement in this out-of-distribution scenario.

\section{Conclusions}
\label{sec:conclusion}

This paper proposes the BEHM-GAN, a method to extend the bandwidth of historical music. The proposed method is based on a generative adversarial network and combines a time-frequency-domain generator with multiple time-domain discriminators. The BEHM-GAN is trained in a self-supervised fashion using lowpass filters to simulate the bandwidth limitation of old recordings. With the intention of strengthening the robustness of our model, we regularize the training by randomizing the cutoff frequency of the filters and perturbing the filtered signal with a small amount of Gaussian white noise. The trained generator is designed to be incorporated as the second step in a music restoration pipeline, where the first step is a deep music denoiser \cite{Moliner2022}. This is, to the best of our knowledge, the first successful work that extends the bandwidth of historical music recordings. 

The proposed method is evaluated using solo piano music, with objective and subjective metrics. As we show in App.~\ref{sec:additional}, the BEHM-GAN can also be applied to other types of music, such as orchestral music or string ensembles. However, this implies retraining the model with specialized data, as our attempts to train the BEHM-GAN with a broader range of music resulted in weaker performance. We leave as future work to study ways to allow the model to have better generalization capabilities without the need for retraining. Another limitation is that our method does not consider all the degradations present in old recordings. While the denoiser does a good job suppressing the additive disturbances and the BEHM-GAN reduces the bandwidth limitation, many other distortion artifacts remain untreated and present in the signal. Further work needs to be done to design a more robust method that addresses these issues.




%

\appendices
\section{Median Scores of the Listening Tests}
Tables \ref{diffs_sim} and \ref{diffs_real} show for each of the listening test sessions, the differences between the median scores of the proposed BEHM-GAN model and the rest of the conditions, respectively. Thus, larger positive values represent worst median scores, relative to the scores received by the proposed method. Asterisks ($\ast$) denote significant differences in a paired t-test, where $\ast$, $\ast \ast$ and $\ast \ast \ast$ respectively indicate p-values $<0.05$, $<0.01$ and $<0.001$.

\begin{table*}[]
\caption{Differences Between the Median Scores of the Listening Test on Synthetic Filtered Recordings}
\label{diffs_sim}
\centering
\begin{tabular}{@{}l|llllll@{}}
\toprule
                          & \multicolumn{6}{c}{\textbf{Synthetic Lowpassed Recordings}}                \\ 
                          & Ex. 1    & Ex. 2    & Ex. 3    & Ex. 4     & Ex. 5    & Ex. 6     \\ \midrule
Low-anchor                & 89 ***   & 100 ***  & 84 ***   & 84.5 ***  & 100 ***  & 78.5 ***  \\
LPF 3kHz           & 56.5 *** & 58.5 *** & 53.5 *** & 50.5 ***  & 54.5 *** & 44.5 ***  \\
AudioUnet                 & 39.5 *** & 56.5 *** & 45.5 *** & 51.5 ***  & 50 ***   & 40.5 ***  \\
SEANet                    & 6        & 29.5 *** & 24.5 *** & 28 ***    & 25 ***   & 21 ***    \\
BEHM-GAN (w/o noise reg.) & 11 *     & 18.5 **  & 29.5 *** & 32.5 ***  & 26.5 *** & 9.5 **    \\
BEHM-GAN (w/o adv.)       & 3.5      & 15.5 *   & 20.5 *** & 8         & 13 *     & 22.5 ***  \\
Reference                 & -15 ***  & 0        & -16 ***  & -15.5 *** & 0        & -21.5 *** \\ \bottomrule
\end{tabular}
\end{table*}
\begin{table*}[]
\caption{Differences Between the Median Scores of the Listening Test on Real Historical Recordings}
\label{diffs_real}
\centering
\begin{tabular}{@{}l|llllll@{}}
\toprule
                     & \multicolumn{6}{c}{\textbf{Real Historical Recordings} }              \\
                     & Ex. 7    & Ex. 8  & Ex. 9    & Ex. 10   & Ex. 11   & Ex. 12   \\ \midrule
Low-anchor           & 77.5 *** & 67 *** & 71.5 *** & 75 ***   & 67.5 *** & 79 ***   \\
Original (denoised)  & 16 *     & 16 *** & 23.5 *** & 24 ***   & 13.5 *   & 26 ***   \\
AudioUnet            & 25.5 *** & 12 **  & 23 ***   & 34 ***   & 20.5 *** & 21 ***   \\
SEANet               & 27.5 *** & -0.5   & 7.5      & 14.5 **  & 4.5 *    & 15.5 *** \\
BEHM-GAN (w/o noise reg.) & 69 ***   & 57 *** & 63.5 *** & 62.5 *** & 57 ***   & 70 ***   \\
BEHM-GAN (w/o adv.)       & 15.5 *** & 20 *** & 31 ***   & 25.5 *** & 11 **    & 9 *   \\  \bottomrule
\end{tabular}
\end{table*}

\section{Experiments with Different Musical Instruments}\label{sec:additional}


 We have also experimented applying our model to other kinds of music having different musical instruments, more precisely string formations and orchestral music. The model has been retrained with a different dataset for each case.   For the first case, 9.5\,h of string ensemble recordings from the MusicNet dataset \cite{thickstun_learning_2017} have been used as training data and, for the orchestral music experiment,  the training data comprised 7\,h of freely-available modern orchestral recordings from The Internet Archive \cite{archive}. 
 
\begin{figure*}[t]
    \centering
    \includegraphics[width=\textwidth]{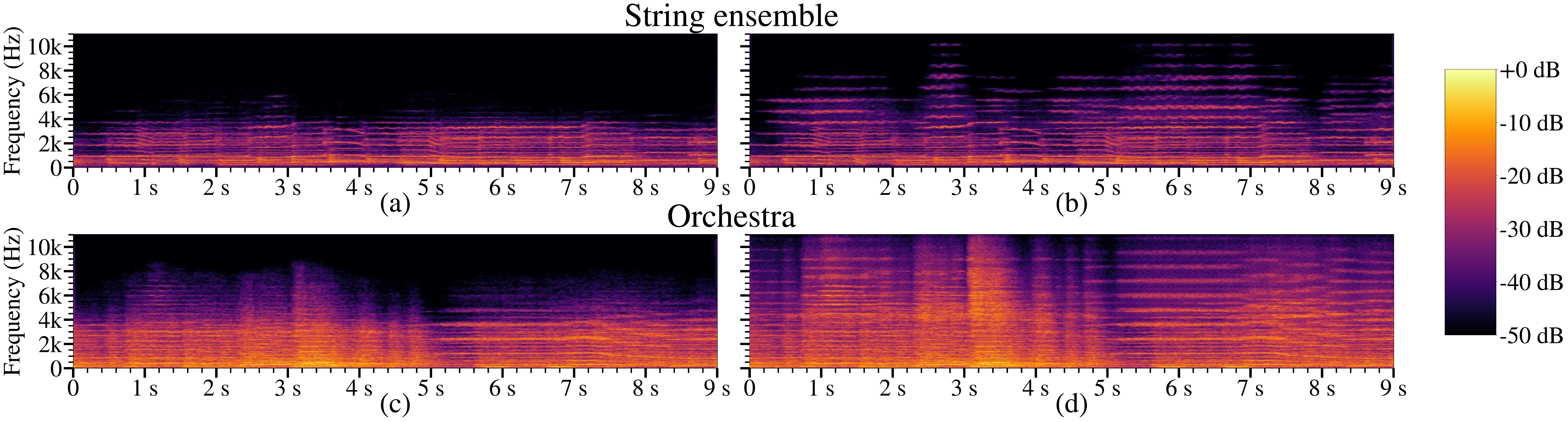}
    \caption{Spectrograms of (a), (c) denoised historical recordings and (b), (d) their bandwidth-extended versions.}
    \label{extra_examples}
\end{figure*}

 Fig.~\ref{extra_examples} shows the spectrogram representations of two original historical recordings and the results after applying the BEHM-GAN. Compared with solo piano recordings, strings and orchestral music recordings have a much richer high-frequency spectra. This implies that the bandwidth extension processing is often more noticeable in these cases. The BEHM-GAN seemed to perform well with string ensemble music. As can be observed in Fig.~\ref{extra_examples}b, the proposed method was able to extend the vibrato sound of a violin. In the case of orchestral music, the model succeeded in enhancing softer passages with strings and winds. However, in this case, we noticed some annoying artifacts with louder percussive instruments, such as drums and cymbals (Fig.~\ref{extra_examples}d). We attribute this weaker performance to the higher variance present in the training data, as orchestral music contains a wide ensemble of different instruments. This variance represents a higher difficulty for the model to generate more robust results. A set of audio examples is available for listening in the companion webpage\footnote{http://research.spa.aalto.fi/publications/papers/ieee-taslp-behm-gan/}.

\newpage
\section*{Acknowledgment}

The authors would like to thank the participants of the listening test. Special thanks go to Mr.~Luis Costa for proofreading the manuscript. Additionally, the authors acknowledge the computational resources provided by the Aalto Science-IT project.

\ifCLASSOPTIONcaptionsoff
  \newpage
\fi



\bibliographystyle{IEEEtran}
\bibliography{IEEEabrv, my_references.bib}
%

%

%

\begin{IEEEbiography}[{\includegraphics[width=1in,height=1.25in,clip,keepaspectratio]{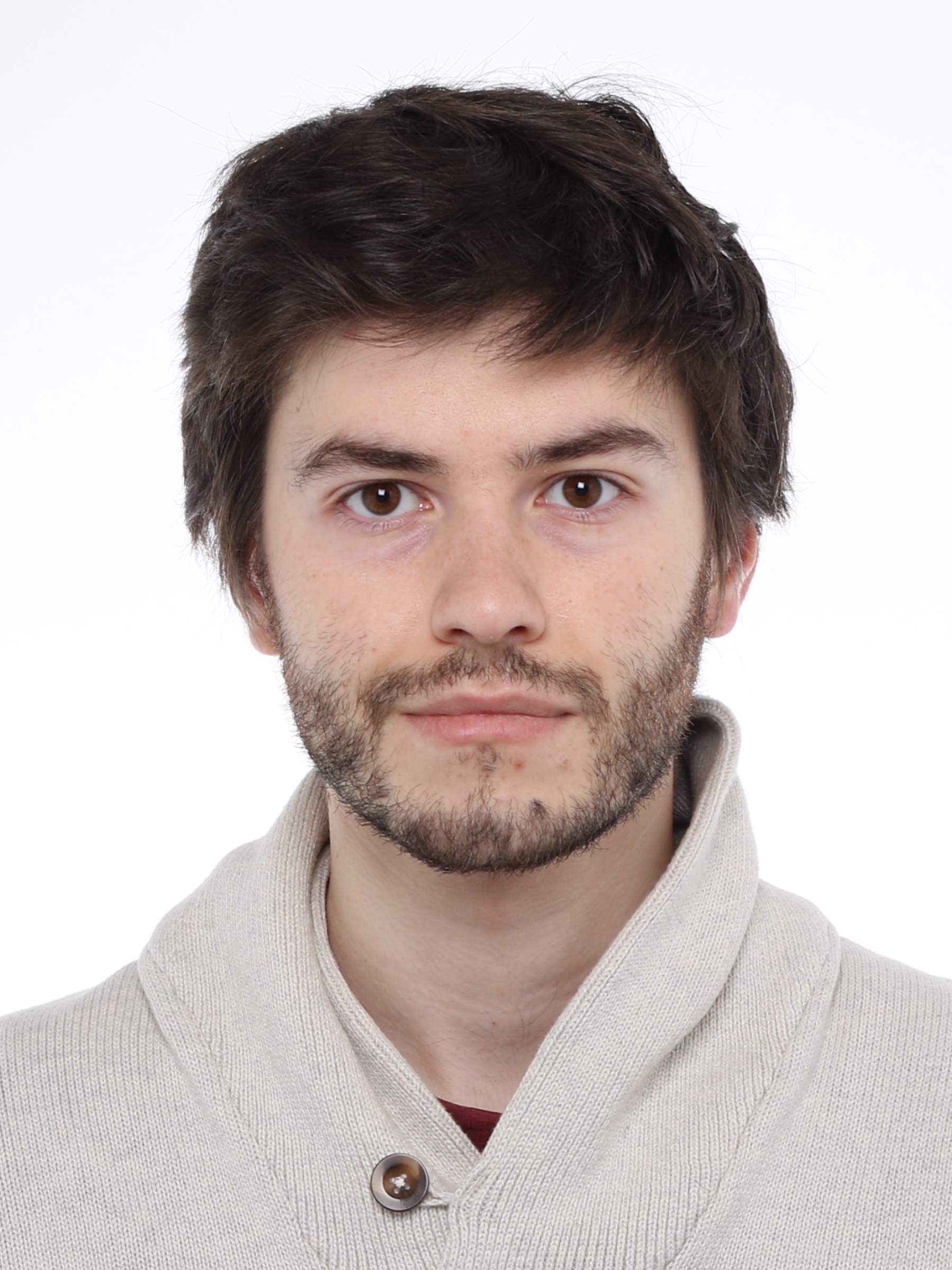}}]{Eloi Moliner}
received his B.Sc. degree in Telecommunications Technologies and Services Engineering from the Polytechnic University of Catalonia, Spain, in 2018 and his M.Sc. degree in Telecommunications Engineering from the same university in 2021. 

He is currently a doctoral candidate in the Acoustics Lab of Aalto University in Espoo, Finland. His research interests include digital audio restoration and audio applications of machine learning.
\end{IEEEbiography}


\begin{IEEEbiography}[{\includegraphics[width=1in,height=1.25in,clip,keepaspectratio]{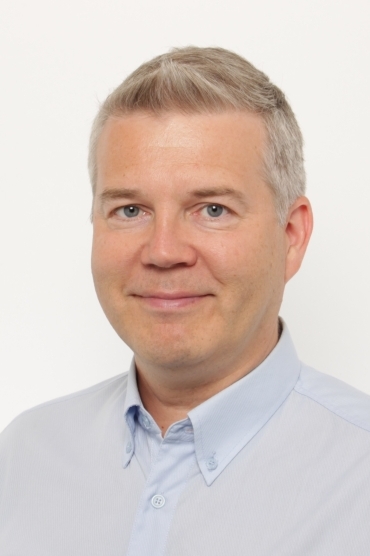}}]{Vesa V\"alim\"aki}
(Fellow, IEEE) received his M.Sc.~and D.Sc.~degrees in electrical engineering from the Helsinki University of Technology (TKK), Espoo, Finland, in 1992 and 1995, respectively.

He was a Postdoctoral Research Fellow at the University of Westminster, London, UK, in 1996. In 1997--2001, he was a Senior Assistant (cf.~Assistant Professor) at TKK. In 2001--2002, he was a Professor of signal processing at the Pori unit of the Tampere University of Technology. In 2008--2009, he was a Visiting Scholar at Stanford University. He is currently a Full Professor of audio signal processing and the Vice Dean for Research in electrical engineering at Aalto University, Espoo, Finland. His research interests are in audio and musical applications of signal processing and machine learning. 

Prof.~V\"alim\"aki is a Fellow of the IEEE and a Fellow of the Audio Engineering Society. In 2007--2013, he was a Member of the Audio and Acoustic Signal Processing Technical Committee of the IEEE Signal Processing Society and is currently an Associate Member. In 2005--2009, he served as an Associate Editor of the {\scshape IEEE Signal Processing Letters} and in 2007--2011, as an Associate Editor of the {\scshape IEEE Transactions on Audio, Speech and Language Processing}. In 2015--2020, he was a Senior Area Editor of the  {\scshape IEEE/ACM Transactions on Audio, Speech and Language Processing}. In 2007, 2015, and 2019, he was a Guest Editor of special issues of the \emph{IEEE Signal Processing Magazine}, and in 2010, of a special issue of the {\scshape IEEE Transactions on Audio, Speech and Language Processing}. Currently, he is the Editor-in-Chief of the {\it Journal of the Audio Engineering Society}.
\end{IEEEbiography}





\end{document}